\documentclass[showpacs,twocolumn,prl]{revtex4}
\usepackage{amsmath}
\usepackage{amsfonts}
\usepackage{graphicx}

\begin{document}

\title{Limit cycle induced frequency locking in
self-sustained current oscillations of superlattices}

\author{Z. Z. Sun}
\author{H. T. He}
\author{J. N. Wang}
\author{Shi-dong Wang}
\author{X. R. Wang}
\email[To whom correspondence should be addressed.
Electronic address: ]{phxwan@ust.hk}
\affiliation{Physics Department,
The Hong Kong University of Science and Technology,
Clear Water Bay, Hong Kong, China}

\date{\today}

\begin{abstract}
The ac response of self-sustained current oscillations
(SSCOs) in GaAs/AlAs superlattices (SLs) is derived based
on the deformation of a limit cycle under an external ac
driving force. Frequency locking into an integer fraction
of the ac frequency is obtained in a periodic response in
which a limit cycle deforms either with or without a
topological change. This frequency locking is robust
against the ac bias because a limit cycle can adjust itself.
The results are verified both numerically and
experimentally, indicating that SSCOs in SLs can be
understood within the framework of the general concepts and
principles of nonlinear physics.
\end{abstract}

\maketitle

Following the early pioneering study\cite{chang,buttiker}
on vertical electron transport in superlattices (SLs),
one of the recent surprising discoveries is self-sustained
current oscillations (SSCOs) under a dc bias\cite{kwok,jwang}.
A large number of experimental and theoretical
studies have focused on different aspects of these
oscillations. It is understood that SSCOs are accompanied
by the motion of boundaries of electric field domains
(EFDs)\cite{kastrup}. A model capable of describing both
the formation of stationary EFDs and SSCOs emerged after
many tedious analyses and numerical calculations\cite
{kastrup}. Our understanding of SSCOs was greatly
advanced through numerical investigations of this model.
As we know, physical system with an intrinsic frequency
may have various possible responses to an external
driving force. For a linear system, such as a simple
pendulum, it will oscillate with the frequency of the
driving force. For a nonlinear system, however, it can
oscillate with an integer multiple of the driving frequency.
For example, a laser light passing through a nonlinear
optical medium may lead to the second and third harmonic
generations. It is also known that many other nonlinear
systems can have frequency locking, in which a system
might oscillate with an integer fraction of the driving
frequency. It is therefore interesting to ask how the
SSCOs observed in SLs response to a combined dc and ac bias.
Numerical solutions on several SL models\cite
{bgb,cao,scholl,ventra} show possible aperiodic oscillations,
either quasiperiodic or chaotic. The experimental evidences
of chaotic behavior were also reported\cite{zhang,luo}.
While most early studies focused on the chaotic behavior of
tunneling current, there were also studies of the periodic
response to an ac bias based on diople EFD model\cite{scholl2}.
However, a generalized explanation with deep insights
into the observed periodic response and other nonlinear
responses in SLs is yet to emerge.

In a recent study\cite{xrw,xrw1}, we found that SSCOs in
a SL correspond to the generation of limit cycles around
an unstable steady-state solution. In the terminology of
nonlinear physics\cite{book}, SSCOs are the manifestation
of one-dimensional attractors-limit cycles. The power of
the limit-cycle concept lies in its simplicity and
universality. An important question one might ask is if
the SSCOs indeed come from limit cycles, what will be
the possible responses of the SSCOs to an extra ac bias?
As it will be demonstrated in this letter, the EFDs model
independent responses of SSCOs in SLs to an extra ac
bias can be readily derived based on limit cycles.
One will see that in periodic responses of SSCOs in SLs the
frequency locking is not only a natural outcome of a limit
cycle, but also very robust against the ac driving force.
Furthermore, we can predict quantitatively the frequency
of a periodic response for a given external ac bias. Thus,
it is beneficial to use the limit cycle to understand SSCOs.
First, by considering a limit cycle as a basic object, we
argue that the limit cycle can have three possible responses
to an extra ac bias:
1) a small deformation without a topological change;
2) a small deformation with a topological change;
3) destruction of the limit cycle.
The first two scenarios lead to the phenomenon of frequency
locking, and the last one gives rise to an aperiodic response.
Second, a widely used drift velocity model is solved
numerically to demonstrate this type of frequency locking and
its robustness. Finally, we present our experimental results.
An excellent agreement between the theory and the experiment
is achieved.

For the sequential electron tunneling in a SL, the phase
space is made up by the bias on each potential barrier,
as explained in References 13 and 14, since the
state of a given SL is fully determined by these biases.
On each point in the phase space, there is a unique vector
which describes the system velocity in the space\cite{book}.
This velocity is determined by the dynamics of the system.
A point with zero velocity is called a fixed point.
An unstable fixed point, as denoted by the cross in Fig. 1,
is such that a small deviation from the fixed point will
drive the system away from the point. However, the system
will stay around the fixed point because of the external
bias constraint. In the case of a SSCO, this local repulsion
and global attraction lead the system to move along a closed
curve, a limit cycle, around the fixed point\cite{xrw,hao}.
Using a two-dimensional case as an example, it is
schematically illustrated in Fig. 1 as a loop in solid line.
Applying an extra ac bias with frequency $\omega_{ac}$, the
velocity field in the phase space changes through the
dynamical equations of the system. If the bias is small, it
can only perturb the velocity slightly, which, in turn,
modifies the limit cycle.

For the periodic response, a limit cycle can change in two
distinct ways. One way is a small deformation of the limit
cycle without a topological change, as shown by the curve
A in Fig. 1. The length of the limit cycle can at most
change a little. The time period for a system to move
along the closed curve once does not change substantially,
since its velocity field in the phase space is controlled
by the system dynamics that is perturbed only slightly by
the extra ac bias. In this case, the system oscillates
with a frequency $\omega$, whose value is not too far from
its intrinsic frequency $\omega_0$.
There is another requirement for the periodic response.
Considering the system starting initially from a point on a
limit cycle, it moves along the limit cycle and returns to
the starting point after a time $T$, giving a frequency
$\omega=2\pi/T$. To have a periodic motion, the external ac
bias should also return to its initial value.
This means that $T$ must be an integer multiple of the
ac-bias period $2\pi/\omega_{ac}$.
Thus, we have $\omega_{ac}/\omega=p=$integer.
In fact, this is a general condition for the periodic motion
of a dynamical system under an ac driving force\cite{book}.
A natural conclusion of this argument is that the limit cycle
makes a small deformation like that of curve A in Fig. 1, when
the ac frequency $\omega_{ac}$ is in the vicinity of $p\omega
_0$, an integer number of the system intrinsic frequency.
 \begin{figure}[hb]
 \begin{center}
\includegraphics[width=7.cm, height=4.5cm]{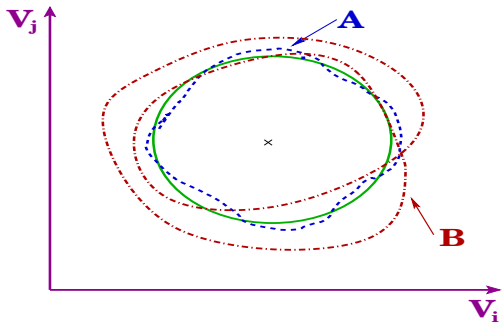}
 \end{center}
\caption{\label{fig1} Schematic drawing of limit cycles
around an unstable fixed point (cross) in a phase plane.
The loop in solid line is the limit cycle in the absence of
an ac bias. Curves A and B are the two possible deformation
of a limit cycle under an ac bias. The system shall oscillate
with a frequency close to its intrinsic one in case A while
it oscillates with half of the intrinsic frequency in case B.}
\end{figure}

The other way is a topological change of the limit cycle in
spite of slight perturbation of the phase velocity field.
This can occur when the system trajectory does not close
itself after moving around the fixed point once. Instead, the
trajectory returns to its starting point after $q$ rounds.
For example, the curve B in Fig. 1 shows a closed curve after
two rounds. In this case, the system oscillates with a
frequency around $\omega_0/2$. It should be pointed out that
this situation can occur only when the dimensionality of
the phase space is larger than two. Combining the periodicity
requirement discussed above, the second scenario occurs only
when $\omega_{ac}/\omega_0$ is in the vicinity of $p/q$
($\neq integer$), where $p$ and $q$ are integers.
Under this type of ac bias, the limit cycle deforms itself
in such a way that it becomes a closed curve after $q$
turns in the phase space, giving $\omega_{ac}/\omega=p$.

Obviously, the periodic response must oscillate with an
integer fraction of the ac frequency if the SSCOs are
indeed due to the generation of limit cycles around an
unstable fixed point. This type of frequency locking does
not depend on a particular model. Since the limit cycle can
deform itself under an ac bias, this periodic response with
an integer fraction of the ac frequency is expected to be
quite robust against $\omega_{ac}$, meaning that $\omega_
{ac}/\omega$ should remain unchanged with a small variation
of $\omega_{ac}$. In order to verify these results, we
numerically solve a widely used discrete drift model
under the combined dc and ac biases in the SSCOs regime.
For a system consisting of $N$ quantum wells under a
bias $U$ between the two end wells, the current flow is
perpendicular to the SL layers. In the sequential tunneling,
charge carriers are in local equilibrium within each well,
so that a chemical potential can be defined locally.
The chemical potential difference between two adjacent
wells is called bias $V$ on the barrier between the two
wells. A current $I_i$ passes through the $i^{th}$ barrier
under a given bias $V_i$. This current may depend on other
parameters, such as doping $N_D$.

Following References 5 and 14, the dynamics of the
system is governed by the discrete Poisson equations
\begin{equation}
\label{poisson}
k(V_i-V_{i-1})=n_i-N_{D}, \qquad i=1,2,\ldots N
\end{equation}
and the current continuity equations
\begin{equation}
\label{charge}
J=k\frac{\partial V_{i}} {\partial t} + I_i
, \qquad i=0,1,2,\ldots N
\end{equation}
where $k$ depends on the SL structure and its dielectric
constant. $n_i$ is the electric charge in the $i^{th}$ well.
In Eq. (\ref{poisson}), the same doping in all wells is
assumed. $I_i$ is, in general, a function of $V_i$ and $n_i$.
It can be shown\cite{sun} that all SSSs are stable if $I_i$
is a function of $V_i$ only. On the other hand, a SSS may be
unstable\cite{kastrup} if one chooses $I_i=n_{i}v(V_{i}),$
where $v$ is a phenomenological drift velocity which is,
for simplicity, assumed to be a function of $V_i$ only.
The constraint equation for $V_i$ is
\begin{equation}
\label{bias}
\sum_{i=0}^{N}V_i=U
\end{equation}
To close the equations, a suitable boundary condition is
needed. It is reasonable to assume a constant $n_0$, $n_0=
\delta N_{D}$, if the carrier density in the emitter is
much larger than those in wells, and its change due to a
tiny tunneling current is negligible.

Previous studies\cite{kastrup,xrw1} have shown that this
model is capable of describing SSCOs with a negative
differential drift velocity. One can obtain a SSCO when
$v(V)=0.0081/[(V/E-1)^{2}+0.01]+0.36/[(V/E-2.35)^{2}+0.18]$,
$N=30$, $U=32.7E$, $N_{D}=0.095kE$, and $\delta=1.001$ are
used\cite{xrw1}. This $v$ has two peaks at $V=E$ and $V=2.35E$.
The region from $V=E$ to $V=1.3E$ exhibits negative
differential velocity. Thus, $E$ can be used as a natural unit
of bias, and $1/v(E)$ as that of the time (the lattice constant
is set to be 1). The intrinsic frequency $\omega_0$ is $0.14
$($v(E)/1$), indicating that the corresponding EFD boundary
oscillates inside about 7 wells. Now we apply an extra ac bias
$V_{ac}\sin(\omega_{ac}t)$ with $V_{ac}=0.327E$ in addition to
the above dc bias. And we have solved numerically the above set
of equations for different $\omega_{ac}$. The current oscillation
frequency can be obtained from the Fourier transformation of
time evolution of the current. The results are plotted in the
$\omega_{ac}/\omega$ vs. $\omega_{ac}/\omega_0$ plane shown by
the solid line in Fig. 2. It has a structure similar to a
devil's staircase. The width of the staircase contains the
information of robustness that the limit cycle can adjust
itself. This robustness depends on both the amplitude and the
frequency of ac bias. The width of the devil's staircase
around $\omega_{ac}/\omega_0=1/q$ decreases with the increase
of $q$ and the results are not displayed for $q\ge 3$.
The dash line is the similar result for $V_{ac}=0.654E$.
The data are offset vertically for a better view.
Clearly, the periodic response around $\omega_{ac}=1.5\omega_
{0}$ disappears, leading to a possible chaotic response.
It shows that the limit cycle is destroyed, and the phase
trajectory, somehow, cannot make a closed curve under this ac bias.
 \begin{figure}[hb]
 \begin{center}
\includegraphics[width=7.cm, height=4.5cm]{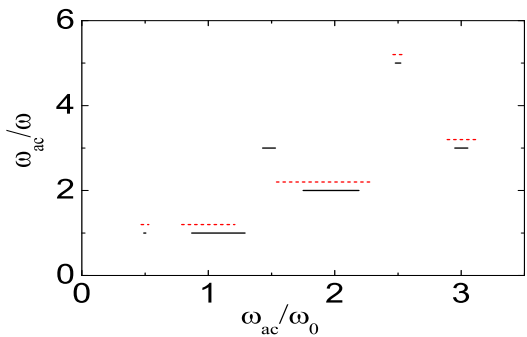}
 \end{center}
\caption{\label{fig2} $\omega_{ac}/\omega$ vs. $\omega_{ac}/
\omega_0$. The devil's staircase type behavior with $\omega_{ac}/
\omega =p$, with integer $p$, shows that the system oscillates
with an integer fraction of ac frequency, and this response is
robust against $\omega_{ac}$. Solid line is for $V_{ac}=0.327E$ and
dashed line is for $V_{ac}=0.654E$. 
Staircases with very narrow widths are not displayed.}
\end{figure}

To verify the above results of frequency locking and its
robustness experimentally, we have measured the response of
SSCOs under an extra ac bias from a GaAs/AlAs SL sample.
The GaAs/AlAs SL sample is grown by molecular beam epitaxy.
It consists of 30 periods of $14$nm GaAs well and 4nm AlAs
barrier and is sandwiched between two $n$+-GaAs layers.
The central 10nm of each GaAs well is doped with Si ($n=2
\times 10^{17}$cm$^{-3}$). The sample is fabricated into $0.2
\times 0.2$mm$^2$ mesas. The SSCOs response is recorded using an
Agilent 54642A oscilloscope. It has been found that SSCOs in a
SL can be induced by changing the sample temperature\cite{jwang}.
In this measurement, the sample temperature is fixed at $95$K and
the dc bias at $0.34$V, which is located within the first plateau
of the time-averaged I-V curve. The inset of Fig. 3 shows the
current oscillation trace without any ac bias. The SSCOs is
clearly demonstrated with a frequency $\omega_{0} = 51.56$KHz.
The oscillation periodicity is indicated by the corresponding
Poincar\'e map (or the first return map)\cite{book}, as shown in
Fig. 3. The Poincar\'e map is derived from the current oscillation
trace by sampling the current trace in a step of $T_0$ (=$2\pi/
\omega_{0}$). Cares have been taken to minimize any possible
artifact in deriving Poincar\'e maps.
 \begin{figure}[hb]
 \begin{center}
\includegraphics[width=7.cm, height=4.5cm]{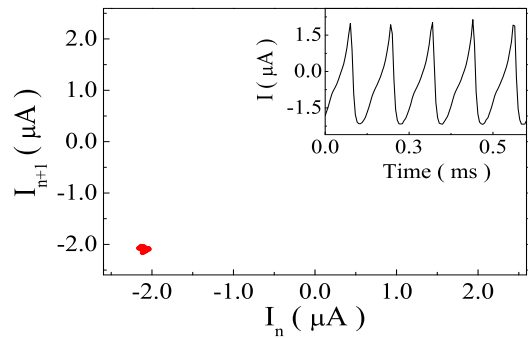}
 \end{center}
\caption{\label{fig3} Poincar\'e map: the points are
($I(nT_0)$, $I(nT_0+T_0)$) for $n=1,2,\ldots$. The accumulation
of all points into a single dot in this map indicates
that the current oscillates periodically with a period of $T_0$.
Inset: time-dependence of tunneling current.
}
\end{figure}
 \begin{figure}[hb]
 \begin{center}
\includegraphics[width=7.cm, height=4.5cm]{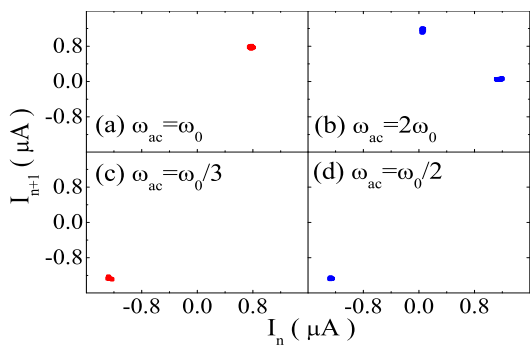}
 \end{center}
\caption{\label{fig4} Poincar\'e maps with sampling step $T_{ac}$
for $\omega_{ac}=\omega_0$ (a); $\omega_{ac}=2\omega_0$ (b);
$\omega_{ac}=\omega_0/3$ (c); $\omega_{ac}=\omega_0/2$ (d).
The number of dots in the maps multiplying $T_{ac}$ are the
corresponding response periods. Thus, the frequencies for (a), (c)
and (d) are all equal to $\omega_{ac}$ while the frequency for (b)
is $\omega_{ac}/2$.
}
\end{figure}

Figure 4 are Poincar\'e maps obtained with an applied extra ac bias.
The ac bias amplitude $V_{ac}$ is set at $66$mV and the driving
frequencies $\omega_{ac}$ are indicated. The Poincar\'e maps are
obtained with sampling steps $T_{ac}$ ($=2\pi/\omega_{ac}$).
As discussed early in this paper, the system exhibits the frequency
locking when $\omega_{ac}$ is set in the vicinity of $p\omega_0$ or
$\omega_0/p$ with $p = integer$. The measured response frequency
$\omega$ is equal to $\omega_{ac}/p$ for $\omega_{ac}=p\omega_0$
or $\omega_{ac}$ for $\omega_{ac}=\omega_0/p$, respectively.
Fig.4 (a) -(d) clearly demonstrate the occurrence of the
frequency locking when $\omega_{ac}=\omega_0,\ 2\omega_0,\
\omega_0/3,$ and $\omega_0/2$, respectively. The two dots in
Fig. 4(b) result from $\omega=\omega_{ac}/2$ while our sampling
step is $T_{ac}$.

In order to demonstrate the robustness of the frequency locking,
we focus on the frequency locking in the vicinity of $\omega_{ac}
=3\omega_0/2$ and $2\omega_0$. By varying the applied ac bias,
the existence of the devil's staircase is shown in Fig. 5, in which
the data for different ac bias amplitude $V_{ac}$ are offset
vertically for clarity. Lines in Fig.5 indicate the frequency
locking range around $\omega_{ac}/\omega_0=1.5$ and $2$ for a
given $V_{ac}$. Clearly, the devils staircase widths, i.e. the
robustness of frequency locking, are strongly dependent on $V_
{ac}$ and $\omega_{ac}$. The locking range for $\omega_{ac}/
\omega_0=2$ is much larger than that for $\omega_{ac}/\omega_0=
1.5$. When $V_{ac}$ is small (=$16$mV) the periodic response
cannot be found around $\omega_{ac}/\omega_0=1.5$.
These results are in good agreement with the theoretical
expectations.
 \begin{figure}[hb]
 \begin{center}
\includegraphics[width=7.cm, height=4.5cm]{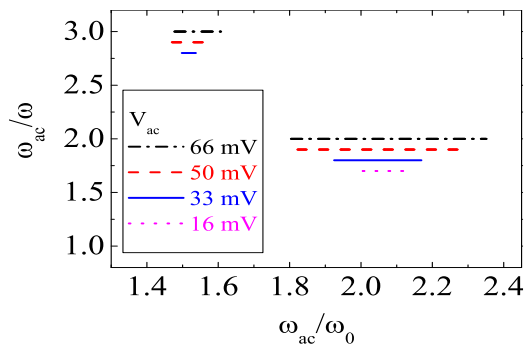}
 \end{center}
\caption{\label{fig5} Experimentally measured periodic response
ranges around $\omega_{ac}/\omega_0=1.5$ and $2$, respectively,
illustrated in $\omega_{ac}/\omega$ vs. $\omega_{ac}/\omega_0$
plots for different amplitudes of ac bias indicated. The data
are offset for clarity.
}
\end{figure}

The above demonstrated quantitative agreements among the theory,
the model calculation, and the experiment clearly indicate
that SSCOs in SLs are indeed originated from the generation of
limit cycles. The unusual frequency locking into a particular
set of an integer fraction of ac frequency is the direct
manifestation of deformation of limit cycles under ac bias.
It is worth to emphasize that based on limit cycles frequency
locking does not depend on the particular structure of the EFD.
Like many other nonlinear dynamical systems, the frequency
locking of SSCOs can be understood within the framework
of the general concepts and principles of nonlinear physics.

In summary, the limit cycle can deform itself in such a way
that it makes $q$ turns in the phase space around an unstable
fixed point when the ac frequency $\omega_{ac}$ is in the
vicinity of $\frac{p}{q}\omega_{0}$ with integers $p$ and $q$.
Thus, a system may oscillate with $\omega_{ac}/p$, an integer
fraction of $\omega_{ac}$, or about $\omega_{0}/q$.
Both of this type of frequency locking and its robustness are
verified by the numerical model calculations and real experiments.
In conclusion, the limit cycle picture of
SSCOs gives a deep insight into the nonlinear properties of SLs.

This work is supported by UGC, Hong Kong, through grants
HKUST6149/00P and HKUST6162/01P. XRW thanks Prof. P. Tong for
useful discussions. We are very grateful to
Prof. Yiping Zeng for providing SL samples.

\end{document}